\pdfoutput=1

\documentclass[usegraphicx,usenatbib]{mn2e}

\usepackage{amssymb,bm,color,charter}
\usepackage{txfonts}
\usepackage{ulem}

\font\tenbg=cmmib10 at 10pt

\def \rvecphi{{\hbox{\tenbg\char'036}}}

\def \rvecmu{{\hbox{\tenbg\char'026}}}
\def \rvecOmega {{\hbox {\tenbg\char'012}}} 

\def\lesssim{\mathrel{\hbox{\rlap{\hbox{\lower4pt\hbox{$\sim$}}}\hbox{$<$}}}}
\def\gtrsim{\mathrel{\hbox{\rlap{\hbox{\lower4pt\hbox{$\sim$}}}\hbox{$>$}}}} 

\title[Kelvin-Helmholtz Instability of the Magnetopause]{Kelvin-Helmholtz Instability of the
Magnetopause of Disc-Accreting Stars}

\author[R.V.E. Lovelace et al.]{R.V.E. Lovelace,$^{1,2}$\thanks{E-mail: RVL1@astro.cornell.edu}
M. M. Romanova,$^1$\thanks{E-mail: romanova@astro.cornell.edu}, 
W.I. Newman,$^3$\thanks{E-mail: win@ucla.edu}\\
$^1$ Department of Astronomy, Cornell University, Ithaca, NY 14853, USA\\
$^2$ Department of Applied and Engineering Physics, Cornell Universisty, Ithaca, NY 14853, USA\\
$^3$ Departments of Earth and Space Sciences,
Physics and Astronomy, and Mathematics, University of
California, Los Angeles, CA 90095}

\begin{document}
\maketitle 
\label{firstpage}

\begin{abstract}

     This work investigates the short wavelength
stability of the magnetopause between  
a rapidly-rotating, supersonic, dense accretion disc and a slowly-rotating
low-density magnetosphere of a magnetized star.
   The magnetopause is  a strong shear layer
with rapid changes in the azimuthal velocity, the density,
and the magnetic field over
a short radial distance and thus the Kelvin-Helmholtz (KH) instability   
may be important.    
     The plasma  dynamics is treated using non-relativistic,
compressible (isentropic)  magnetohydrodynamics. 
It is necessary to include the displacement current  in order that
plasma wave velocities remain less than the speed of light.
    We focus mainly on the case of a star with an aligned
dipole magnetic field so that the magnetic field
is axial in the disc midplane  and
perpendicular to the disc flow velocity.   
  However, we also give results for cases where
the magnetic field is at an arbitrary angle to the flow
velocity.   
      For the aligned dipole case the magnetopause is most  unstable     
for KH waves propagating  in the azimuthal
direction perpendicular to the magnetic
field which tends to stabilize waves propagating parallel to it.  
  The wave phase velocity is that of the disc matter.
    A quasi-linear theory of the saturation of the instability
leads to a wavenumber ($k$) power spectrum $\propto k^{-1}$
of the density and temperature fluctuations of the
magnetopause, and it gives the
 mass accretion and
angular momentum inflow rates across the
magnetopause.    
    For self-consistent conditions
this mass accretion rate will be equal to the 
disc accretion rate at large distances from
the magnetopause.

\end{abstract}

\begin{keywords}  accretion, accretion discs ---  stars: neutron
--- X-rays: binaries --- magnetohydrodynamics --- Instabilities --- Waves
\end{keywords}

\section{Introduction}

     This work investigates the short wavelength
stability of the interface or magnetopause between  
a rapidly rotating accretion disc and the slowly-rotating,
low-density magnetosphere of a magnetized star.
    The nature of the magnetopause is sketched
in Figure 1.        
    The rotating  disc matter 
is ``held off'' by the star's  magnetosphere
where the magnetic field is strong and
the density is small.  The disc matter rotates
at approximately the Keplerian velocity 
which is typically much
larger than the velocity of the magnetospheric
plasma which corotates  with the angular
velocity of the star.   
 Thus the interface involves 
a strong shear layer as  sketched in the bottom part
of Figure 1.   
     Understanding the instabilities of
the magnetopause is important for 
understanding both the transport of matter
and angular momentum towards the star {\it and} the temporal
variability of the sources (van der Klis 2006).

    The magnetohydrodynamic (MHD) stability of configurations
such as in Figure 1    
was investigated earlier by Li \& Narayan (2003)
assuming incompressible flow and perturbations
independent of $z$, but with no restrictions
on the azimuthal wavelength.  
      They found both
long-wavelength (i.e., $\lesssim r$) Rayleigh-Taylor (RT)
 and short wavelength ($\ll r$) Kelvin-Helmholtz (KH) 
instabilities for different conditions of the shear layer.  
   Earlier, Burnard, Lea, \& Arons (1983) studied
the short wavelength KH instability of the magnetopause
of a spherically accreting rotating magnetized star.   
    Recently Tsang \& Lai (2009) have studied the long
wavelength RT stability of the sharp interface including
the compressibility of the media.
    The MHD stability of the magnetopause for cases
where the shear layer has appreciable radial
width was studied by
Lovelace \& Romanova ((2007) and Lovelace,
Turner, \& Romanova (2009)   for compressible,
non-barotropic perturbations independent of $z$
but no restrictions on the  azimuthal wavelength.
   They found a resonant long-wavelength  Rossby Wave Instability (RWI;
Lovelace et al. 1999) which may contribute to
the observed twin kilo-Hertz quasi-periodic oscillations of
low-mass X-ray binaries
(van der Klis 2006).

    Here we consider a thin magnetopause and
short wavelengths where the   KH instability
is expected to be important.  
    There is a vast literature on
the magnetized KH instability in 
different astrophysical and space applications,  to
the magnetopause of rotating planets in
the solar wind  (e.g., Miura \& 
Pritchett 1982; Roy-Choudhury \&
Lovelace 1986; Faganello, Califano, \& Pegoraro 2008),
to the stability of astrophysical jets (e.g., Hardee 2007; Osmanov
et al. 2008),  to the stability of interfaces of molecular/atomic 
clouds in the interstellar medium (e.g., Hunter, Whitaker, \& 
Lovelace 1998), and to the interface of an unmagnetized
dense plasma blob falling through the strong
magnetic field of a neutron star (Arons \& Lea 1980).

    Section 2 of the paper gives the basic equations where
the fluid motion is assumed non-relativistic but the 
displacement current is retained in Maxwell's  
equations in order to keep the wave speeds less
than the light speed.  
This section also describes
the assumed equilibrium, the waves in the 
disc plasma, and the MHD waves
in the magnetospheric plasma.    
   The disc flow speed is
much larger than the disc sound speed.
Section 3   obtains a
dispersion relation for the Kelvin-Helmholtz modes
and develops an approximates solution for cases where
the magnetopsheric density ($\rho_2$)  is much less than the
disc density ($\rho_1$).    
   Section 4 discusses the non-linear saturation of the KH
modes and develops    a quasi-linear theory for the 
mass accretion rate across the magnetopause.
   Section 5 discusses briefly the case where the
magnetic field is at an arbitrary angle relative to
flow velocity in the disc.   
   Section 5 gives the conclusions of this work.

\begin{figure}
\centering
\includegraphics[scale=0.55]{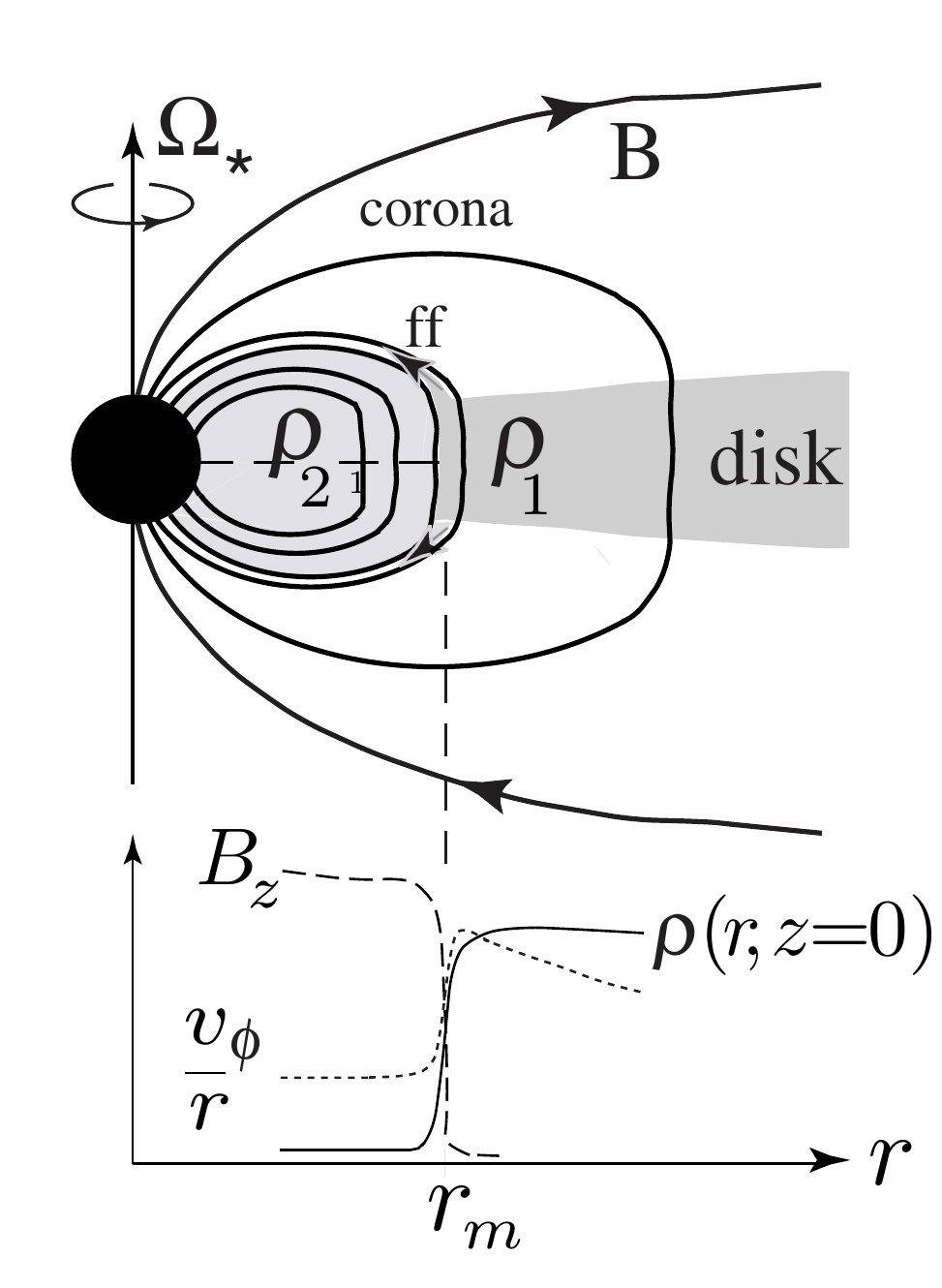}
\caption{Sketch of the inner region of a disc around a rotating
magnetized star suggested by MHD simulations
of Romanova et al. (2008).      
   Inside of $r_m$ the
plasma flows along the magnetic field lines to the surface
of the star in what is termed a funnel flow (denoted ff in the figure).
The bottom part of the figure shows the
midplane profiles of the the density $\rho$, magnetic field
$B_z$, and azimuthal velocity $v_\phi$. }
\end{figure}

\section{Theory}

    The rotating disc matter is termed
region 1, while the magnetosphere
matter is termed region 2.
    In the magnetosphere the magnetic
field     ${\bf B}$ is 
significant whereas in the external
medium  ${\bf B}$ is  negligible as
suggested by MHD simulations (Romanova, Kulkarni,
\& Lovelace 2008; Kulkarni \& Lovelace 2008).
  Depending on the region,   the flow is described by 
non-relativistic 
hydrodynamic equations or by
MHD equations,
\begin{equation}
\rho { d {\bf v} \over dt}= 
-{\bf \nabla} p 
+{1\over c} {\bf J \times B}+\rho {\bf g}~, ~~~
{\partial \rho \over \partial t}
+{\bf \nabla }\cdot (\rho {\bf v})=0~,
\end{equation}
where ${\bf g} = -\nabla \Phi$ is the 
gravitational acceleration due to the star and $\Phi$
is the gravitational potential.
   In addition we have
Maxwell's equations including
the displacement current,
\begin{equation}
{\bf \nabla \times B} = 
{4 \pi \over c} {\bf J}+{1\over c}
{\partial {\bf E} \over \partial t}~,\quad
{\bf \nabla \times E } = 
-\frac{1}{c}\frac{\partial {\bf B}}{\partial t}~,
\end{equation}
and Ohm's law for infinite conductivity,
\begin{equation}
0 = {\bf E} 
+ {\bf v \times B}/c~,
\end{equation}
where ${\bf v}$ is the flow velocity,
${\bf B}$ the magnetic field,
$p$  the plasma pressure, and $\nabla\cdot {\bf B}=0.$
   We assume isentropic
flow with $\gamma=5/3$ in both media:  $p=\kappa_1\rho^\gamma$ in
the external medium (region 1) and $p=\kappa_2\rho^{\gamma}$ in
the magnetospheric plasma (region 2).

\begin{figure}
\centering
\includegraphics[scale=0.55]{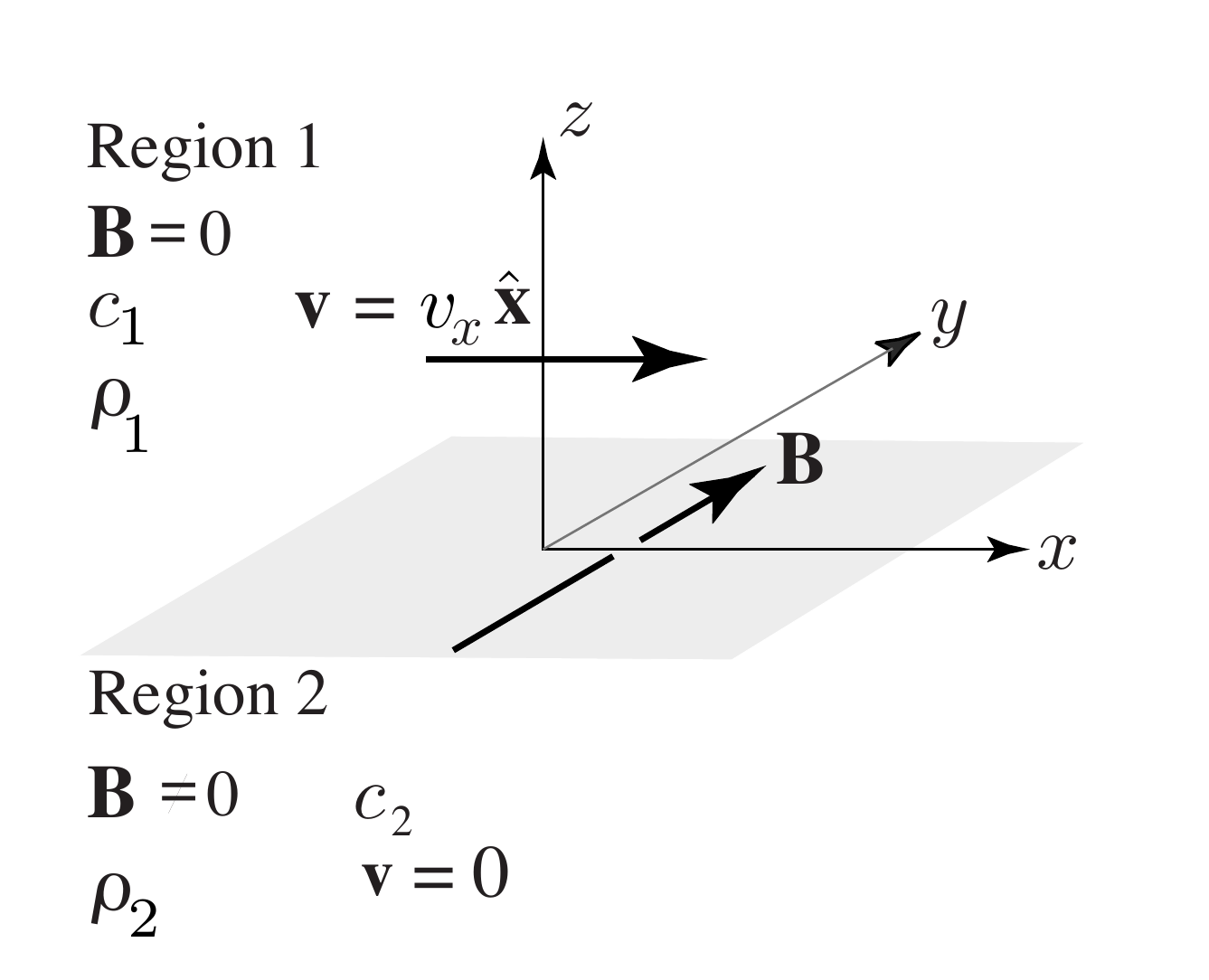}
\caption{Geometry of a small patch of the
interface between the star's magnetosphere
and the external disc flow.
}
\end{figure}

     One can eliminate ${\bf E}$ and ${\bf J}$
from the equations to obtain
\begin{equation}
\frac{\partial {\bf B}}{\partial t}=
{\bf \nabla \times (v \times B) }~,
\end{equation}
and
\begin{equation}
\rho { d {\bf v} \over dt}= 
-{\bf \nabla} \left(p +{{\bf B}^2 \over 8 \pi}\right)
+{\bf B \cdot}{\bf \nabla}\left({{\bf B}\over 4\pi}\right)
-{1 \over 4\pi c^2}{\bf B}\times {\partial \over \partial t}
({\bf v \times B})+\rho {\bf g}~.
\end{equation}
The less familiar last term of this equation arises
from the displacement current in equation (2).  It is 
required in order to have the wave speeds less
than the velocity of light (Jackson 1975).

\subsection{Equilibrium Flow}

    We consider an axisymmetric time-independent 
equilibrium plasma flow.
    The flow velocity
${\bf v} = v_\phi(r) \hat{\rvecphi~}=r\Omega_\phi(r) \hat{\rvecphi~}$.  
   That is, the accretion velocity $v_r$ and the vertical
velocity $u_z$  are assumed negligible compared with $v_\phi$.
     Initially, we use an inertial cylindrical
$(r,\phi,z)$ coordinate system.
      The equilibrium magnetic field
is  ${\bf B}=B(r)\hat{\bf z}$.   
    In \S 5 we discuss the general case where
the magnetic field has a general orientation with
respect to the flow velocity in the disc.    
The equilibrium flow satisfies $-\rho g=
-d[p+B^2/(8\pi)]/dr$.  Here, $\rho$ is the density, $p$ the pressure, and
$\Omega_K$ the Keplerian angular rotation rate of a single particle.
  We assume a pseudo-Newtonian potential where
 $\Omega_K^2= GM/[r(r-r_S)^2]$ with $r_S=2 G M/c^2$ the
 Schwarzschild radius of the star. 
     The effective radial gravitational acceleration is $g=-r (\Omega_K^2-\Omega_\phi^2)$.

    We consider the case where the plasma properties
undergo  a rapid change at the 
magnetopause radius $r_m$ as sketched in the bottom
of Figure 1.   The change in values is assumed to
occur over a distance $\Delta r_m \ll r_m$.
      Inside of $r_m$ the magnetic
field is $B$, the density  is $\rho_2$, the sound
speed is $c_2$, and
the flow velocity is $r\Omega_* \hat{\rvecphi~}$, where
$\Omega_*$ is the angular rotation rate of the star.    
    Outside of $r_m$ the magnetic field is negligible,
the density is $\rho_1$, the sound speed
is $c_1$, and the flow velocity
is  close to Keplerian,   $v_K\hat{\rvecphi~}$.
    We consider the case suggested by 3D MHD
simulations (Romanova, Kulkarni, \& Lovelace 2008;
Kulkarni \& Romanova 2008)
where    $\rho_2 \ll \rho_1$ and
$c_2 \sim c_1$.
    For such conditions the force balance at
$r_m$ requires that $B^2/8\pi \approx p_1=
\rho_1 c_1^2/\gamma$.

   We focus our attention on the stability of short wavelengths
$\lambdabar =\lambda/2\pi  \ll  r$.   
     This allows us
to consider the stability of  a small patch of the magnetopause
separating the two plasmas as indicated in Figure 2.    
The change of coordinates,  from cylindrical to Cartesian, is $(r\rightarrow z, ~r\phi \rightarrow x,~ z\rightarrow y)$. 
     For the considered short wavelengths
the gravitational force can be neglected in equation (5).
That is, we consider the Kelvin-Helmholtz 
rather than the Rayleigh-Taylor modes.
     We choose a reference frame moving with
the magnetospheric plasma.   
  Consequently the external disc plasma moves with
a velocity ${\bf v}=v_x\hat{\bf x}$ with $v_x 
= (v_K -r\Omega_*)_{r=r_m}$.

\begin{figure}
\centering
\includegraphics[scale=0.35]{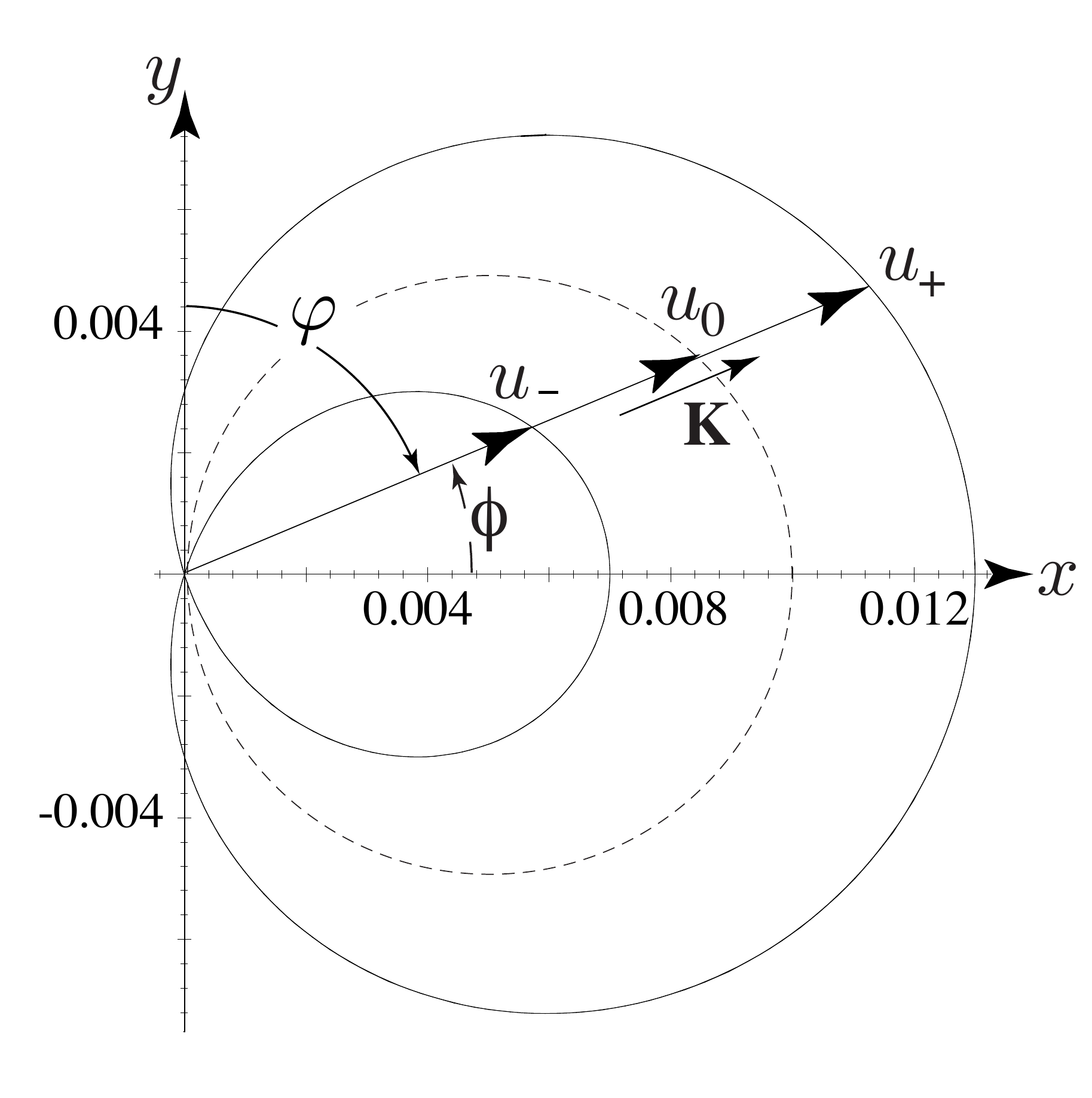}
\caption{Phase velocities $u_\pm
=\omega_\pm/|{\bf K}|$ (fast and slow) of sound
waves in the external medium viewed from
a reference frame where the magnetosphere
is at rest.
  For the case shown the external medium moves
 with  velocity $v_x=0.01$ and
the sound speed is $c_1=0.003$ with
speeds in units of $c$.  The fast/slow phase velocities
go through the $x-$axis at $v_x \pm c_1$.
}
\end{figure}

\subsection{Waves in the External Medium}

       In region 1 there is no magnetic field so that
the small amplitude
waves $\sim \exp(i{\bf K \cdot r}-i\omega t)$
in the rest frame of the medium consist of {\it both}
sound waves and non-propagating disturbances
such as variations in the vorticity 
(Landau \& Lifshitz 1987).
   The dispersion relation for the sound waves is
$\omega^2 = {\bf K}^2 c_1^2$,
where $\omega$ is the angular frequency, 
${\bf K}=(K_x,K_y,K_z)$ is the wavevector,
$c_1=
(\partial p/\partial \rho)_S^{1/2}$ is the sound speed
 in the external medium.
   In the reference frame we use, the external medium is moving
with uniform velocity $v_x$ so that the dispersion
relation in this reference frame is
\begin{equation}
(\omega -K_x v_x)^2= {\bf K}^2 c_1^2~.
\end{equation}
which gives $\omega_\pm = 
K_x v_x \pm |{\bf K}|c_{s1}$.   
   For the non-propagating disturbances,
$\omega_0 = K_x v_x$.   
   Figure 3 shows a polar plot of the phase velocities
$u_\pm =\omega_\pm /|{\bf K}|\geq 0$ and
$u_0=\omega_0/|{\bf K}|$ of these waves
as a function of $\phi$ which is the angle
between ${\bf K}$ and the $x-$axis.

\begin{figure}
\centering
\includegraphics[scale=0.45]{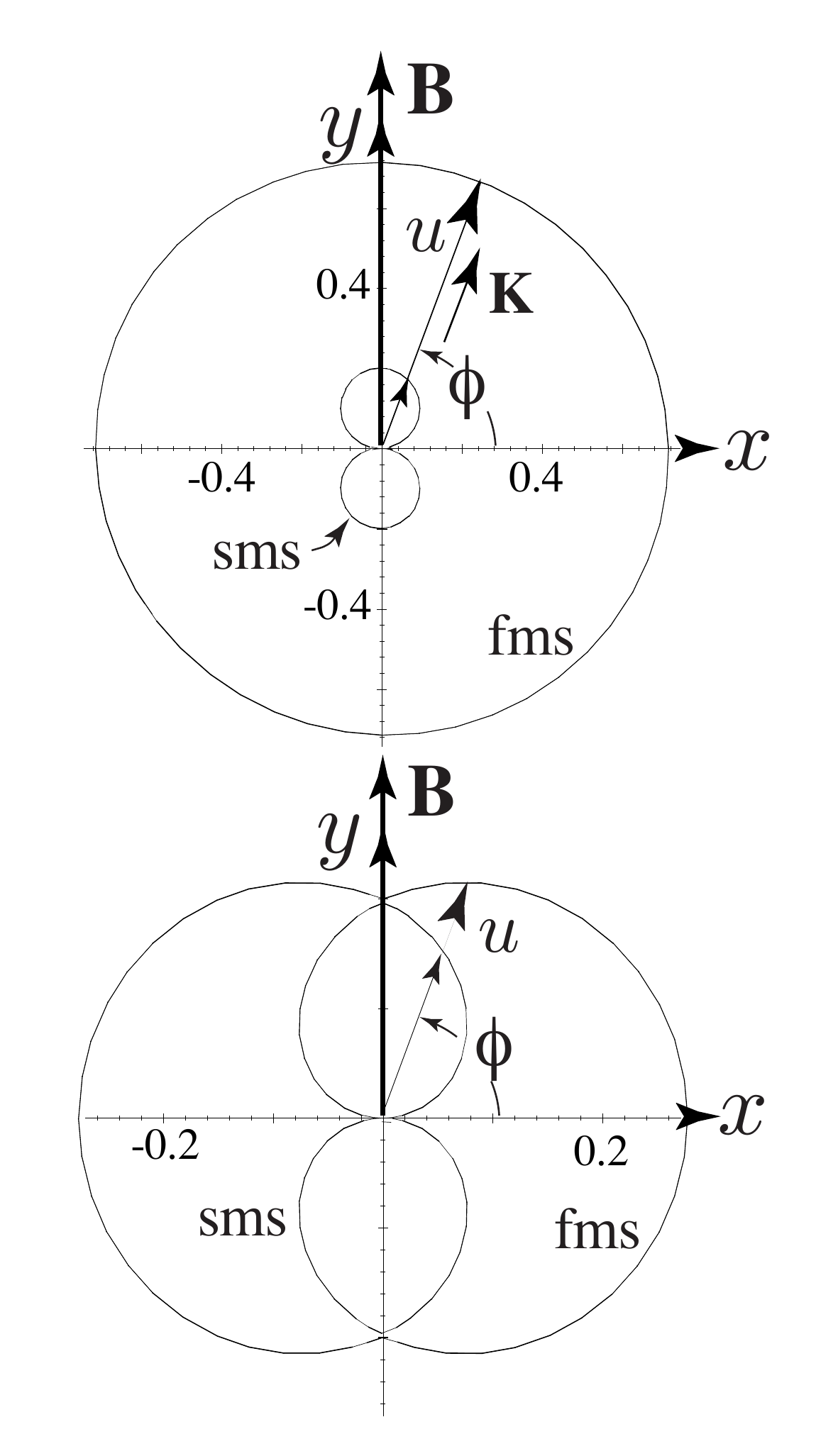}
\caption{Phase velocities $u$ of slow magnetosonic
(sms) and fast magnetosonic waves (fms) in
the magnetosphere.
    The magnetic field direction is assumed
 to be along the $y-$axis.
   In the top panel of the figure
has $v_A=1$ and $c_{m}=0.2$,
while in the bottom has $v_A=0.2$ and $c_{m}=0.2$,
where the speeds are in units of $c$.
}
\end{figure}

\subsection{MHD Modes in the Magnetosphere}

The dispersion relation for the 
two compressible MHD modes including
the displacement current is
\begin{equation}
\omega^4
-\omega^2\left[{\bf{K}}^2 (\tilde{c}_s^2+
\tilde{{v}}_A^2)+({\bf K \cdot v}_A)^2
\tilde{c}_s^2/c^2
\right]
+{\bf K}^2  \tilde{c}_s^2 ({\bf K \cdot v}_A)^2
 =0~,
\end{equation}
where 
\begin{equation}
\tilde{c}_{s}^2 \equiv {c_{s}^2 \over 1+{{\bf
v}}_A^2/c^2}~, ~{\rm where}~~~ 
c_{s} \equiv \left({\partial p 
\over \partial \rho} \right)_S^{1/2}~,
\end{equation}
(e.g., Jackson 1975).  
Here,
$\omega$ is the angular frequency of the
wave and $\bf K$ is the wavevector.
   Note that the bold-faced
${\bf v}_A$'s in equation (9) do not have over-tildes
and can be larger than $c$.
    
  The magnetic field in the magnetosphere
is strong and the plasma density low so that we expect to
have
$\tilde{v}_A^2 \gg \tilde{c}_s^2$ or equivalently $v_A^2 \gg c_s^2  $.  
   In this
limit the slow magnetosonic 
wave branch of equation (7) 
has $\omega^2_{sm} \approx {\bf K}^2 c_{s}^2
\cos^2(\varphi)$, where $\varphi$ is
the angle between ${\bf K}$ and ${\bf B}$.  
   For this wave  the velocity perturbations
are parallel to ${\bf B}$ with the result
that the wave speed is independent of ${\bf B}$.
 The other, fast magnetosonic
branch has $\omega^2_{fm} \approx {\bf K}^2 \hat{
v}_A^2$ with $ \omega_{fm}^2\gg \omega_{sm}^2 $.
   This wave has velocity perturbations in
the plane formed by ${\bf K}$ and ${\bf B}$.  
    Both waves give density variations.   There
is also a non-propagating wave analogous to that
in region 1.    
   Figure 4 shows a polar plot of the phase velocities of
the waves obtained from equation (7).

    The dispersion relation for the
shear Alfv\'en wave  including
the displacement current is
\begin{equation}
 \omega^2 = ({\bf K}{\bf \cdot} \tilde{\bf v}_A)^2~,
\end{equation}
where
\begin{equation}
\tilde{ \bf v}_A \equiv  { {\bf v}_A \over
(1+ {\bf v}_A^2/c^2)^{1/2}}~,
\end{equation}
where $ {\bf v}_A \equiv {\bf B}/(4\pi\rho)^{1/2}$ is the
usual, non-relativistic Alfv\'en velocity
which may be larger than $c$ (e.g., Jackson 1975).
   The velocity perturbation of this wave and
the field perturbation $\delta {\bf B}$
are perpendicular to both ${\bf K}$ and ${\bf B}$.
Thus the wave does not change the fluid density.

\begin{figure}
\centering
\includegraphics[scale=0.65]{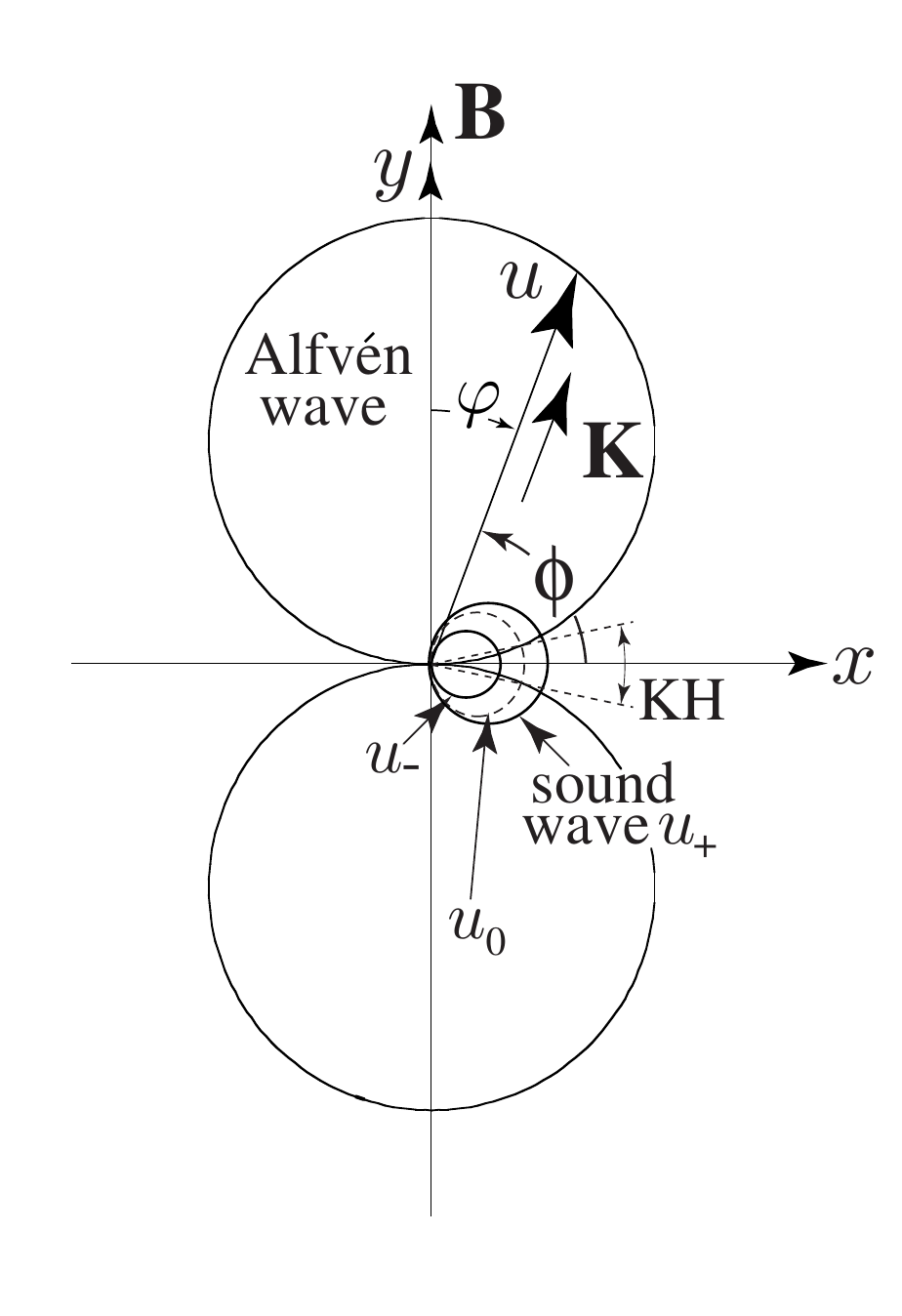}
\caption{Phase velocities $u$ of the Alfv\'en
wave in the magnetosphere as well as the fast
and slow sound waves $u_\pm$ in the external
plasma.   In this figure, $c_{s1}=c_{s2}=0.0391$,
$\tilde{v}_A =0.805$, and $v_x=0.167$ all in units
of $c$.   The dashed circle is $v_x \sin(\varphi)$
(for $0 \leq\varphi \leq \pi$) and
the dashed lines delimit the wavevector region of
the Kelvin-Helmholtz instability discussed
in \S 3.2.
}
\end{figure}

\section{Linearized Equations for  the KH Modes}

   The geometry of a small
patch of  the interface between
the external medium and the magnetosphere is shown 
in Figure 2.  
   The interface
is in pressure equilbrium so that
$p+ {\bf B}^2/8\pi$ is continuous 
across it.

   Perturbations of all scalar quantities
have the form $f(z)\exp(i{\bf k \cdot r}-i\omega t)$
for all $z$, where ${\bf r}=(x,y,0)$,
 ${\bf k} = (k_x,k_y, 0)$ is the wavevector
in the plane of the interface, 
and the $z-$dependence remains to be
determined.
     The linearized continuity equation is
\begin{equation}
-i\Delta \omega \delta \rho +
i\rho {\bf k}\cdot \delta {\bf v}
+{\partial \over \partial z}
\left( \rho \delta {v}_z\right) =0~,
\end{equation}
where $\delta {\bf v}=(\delta v_x,\delta v_y,0)$ and 
$\Delta \omega
\equiv
\omega - k_x v_x$ is the Doppler shifted wave frequency in
the frame comoving with the external medium.
   The step-function dependences
of $\rho(z)$, $v_x(z)$, $ B_x(z),$ and
$B_y(z)$, gives rise to various
delta-function terms in the linearized
equations.  
   For example, from the top side (region 1, $z>0$)
the vertical displacement  of the interface
is $\delta z_1$ and the velocity at $z=\epsilon^+$ is
$$
\delta v_{z1}=\left({\partial \over \partial t}
+v_x{\partial \over \partial x}\right) \delta z_1=-i\Delta
\omega \delta z_1~,
$$
while from the bottom side (region 2)
it is
$$
\delta v_{z2} ={\partial \over \partial t}
\delta z_2 = -i\omega \delta z_2~.
$$
We must have $\delta z_1 = \delta z_2$ 
so that
\begin{equation}
{\delta v_z(z=\epsilon^+) \over \omega - k_xv_x(\epsilon^+)}
={\delta v_z(z=\epsilon^-) \over \omega}~,
\end{equation}
(Chandrasekhar 1961), where $\epsilon^\pm$ denotes
an arbitrarily small positive or negative quantity.
  Thus $\delta v_z$ is discontinuous
across the interface.
    Considering perturbations
proportional to $\exp(ik_{z 1,2} z)$,
equation (11) can be written as 
\begin{equation}
\delta \rho= {\rho(z){\bf K \cdot}\delta{\bf v} \over
\omega -k_x v_x(z)} - {i \delta v_z \over \omega
-k_x v_x(z)}{\partial \rho \over \partial z}~.
\end{equation}
Here, ${\bf K}=(k_x,~k_y,~k_z)$ is the three-dimensional
wavenumber which also comes up later.
The  second term of this equation gives
a delta function dependence, $-\delta z_1 
(\rho_1-\rho_2)\delta(z) $.

   Linearization of the Euler equation (5) gives
\begin{eqnarray}
-iW \rho \delta v_x = - ik_x \delta \hat{p}
+i({\bf k \cdot B}) {\delta B_x\over 4\pi}
-i\omega B_x{{\bf B}\cdot \delta {\bf v}\over 4\pi c^2}
\nonumber\\
+i\Delta \omega \rho {d v_x \over dz}\delta z 
+{1 \over 4\pi}  {dB_x \over dz}\delta B_z~,
\end{eqnarray}
\begin{eqnarray}
-iW \rho \delta v_y = - ik_y \delta \hat{p}
+i({\bf k \cdot B}) {\delta B_y\over 4\pi}
-i\omega B_y{{\bf B}\cdot \delta {\bf v}\over 4\pi c^2}
\nonumber\\
+{1 \over 4\pi}  {dB_x \over dz}\delta B_z~,
\end{eqnarray}
\begin{equation}
-iW \rho \delta v_z = - {\partial \over
\partial z}\delta\hat{p}
+i({\bf k \cdot B}) {\delta B_z\over 4\pi}~,
\end{equation}
where 
$$
W\equiv \Delta \omega +\omega {v_A^2 \over c^2 }~,
$$
 and
$$
\delta \hat{p}\equiv \delta p +{1\over 4\pi}(B_x\delta B_x
+B_y\delta B_y)~.
$$
The terms $\propto 1/c^2$ are due to the
displacement current and they must in general
be retained.
   Recall that ${\bf B}=B_x\hat{\bf x}+B_y\hat{\bf y}$
so that ${\bf K \cdot B}={\bf k \cdot B}$.

     Linearization of the induction
 equation (4) for the magnetic field
gives
\begin{equation}
\delta B_x={B_x}{\delta \rho \over \rho}
-{{\bf k \cdot B} \over \Delta \omega} \delta v_x
-{{\bf k \cdot B} \over \Delta \omega} {dv_x \over dz}
\delta z -{dB_x\over dz} \delta z +
{B_x \over \rho}{d\rho \over dz}\delta z~,
\end{equation}
\begin{equation}
\delta B_y={B_y}{\delta \rho \over \rho}
-{{\bf k \cdot B} \over \Delta \omega} \delta v_y
 -{dB_y\over dz} \delta z +
{B_y \over \rho}{d\rho \over dz}\delta z~,
\end{equation}
\begin{equation}
\delta B_z =i({\bf k \cdot  B})\delta z = 
-{{\bf k \cdot B} \over \Delta \omega} \delta v_z~.
\end{equation}
At the interface the terms
involving $\partial(v_x, \rho, B_x, B_y)/\partial z$
give delta functions $\sim \delta(z)$.
Away from the interface these terms vanish.

   Away from the interface we can combine 
equations (14) and (15) and (17) and (18) to obtain
\begin{equation}
\rho {\bf B}\cdot \delta {\bf v}=
{\Delta \omega \over F} ({\bf k \cdot B}) 
\big(\delta \hat{p}- {v}_A^2 \delta \rho\big)~,
\end{equation}
where 
$$
F \equiv \Delta \omega^2 -({\bf k \cdot v}_A)^2~,
$$
and ${\bf v}_A \equiv {\bf B}/\sqrt{4\pi \rho}$.
Note ${\bf K \cdot \delta v}=
\Delta \omega(\delta \rho/\rho)$ and that
${\bf K \cdot \delta B} =0$.

    As mentioned we use the reference
frame comoving with the magnetosphere
so that in region 2    
 $\Delta \omega=\omega$.
   From equations (14) and (15) we have
\begin{equation}
\omega\left(1+{v_A^2\over c^2}\right)\rho {\bf K \cdot \delta v}=
{\bf K}^2 \delta \hat{p} +{ \omega  \over 4\pi c^2}
({\bf k \cdot B}) {\bf B \cdot \delta v}
~.
\end{equation}
  From equations (17) and (18) we
have
\begin{equation}
{\bf B} \cdot \delta{\bf B}={\bf B}^2 
{\delta \rho \over \rho}
-{{\bf k \cdot B}\over \Delta \omega}
{\bf B}\cdot \delta {\bf v}~.
\end{equation}
  Using the above relation between $\delta p$
and $\delta \hat{p}$, and 
the relation for
isentropic perturbations 
$\delta p = c_s^2 \delta \rho$, we obtain
\begin{equation}
\delta \hat{p} = \left(c_s^2+{\bf v}_A^2 -
{({\bf k \cdot v}_A)^2 c_s^2 \over
\Delta \omega^2}\right)\delta \rho~.
\end{equation}
Equations (21) - (23) can be readily combined
to give the dispersion relation for the
magnetospheric modes.

    Away from the interface  equations
(11) and (15) give
\begin{equation}
{\partial^2 \over \partial z^2}\delta \rho =
 - k_z^2 \delta \rho~.
\end{equation}
   For the external disc 
medium, $z>0$, we must have $\delta \rho \propto
\exp(ik_{z1}z)$ with $\Im(k_{z1}) >0$
so that the perturbation decays as $z$ increases.
    In this region we have
\begin{equation}
k_{z1}^2= - {\bf k}^2 +{(\omega - k_xv_x)^2\over c_{s1}^2}~,
\end{equation}
which is identical to equation (6) with
${\bf K}^2 ={\bf k}^2+k_z^2$ as it should be.
Dividing this equation by ${\bf k}^2$ and taking
the square root gives
\begin{equation}
\tilde{k}_{z1} (u)=\pm \left(  {[u-\sin(\varphi)v_x]^2 
\over c_{s1}^2}-1\right)^{1/2}~,
\end{equation}
where $\tilde{k}_z \equiv k_z/|{\bf k}|$,  $\varphi$
is the angle between $\bf k$ and $\bf B$, and
$u\equiv \omega/|{\bf k}|$ is the phase
velocity of the perturbation which is complex with
a positive imaginary part for an unstable perturbation.
   The choice of the sign in this expression is
determined by the condition $\Im(\tilde{k}_{z1} >0$.

  For the magnetospheric
plasma, $z<0$, we  must have $\delta \rho \propto
\exp(ik_{z2} z)$ with $\Im(k_{z2})<0$ so
that the perturbation decays as $-z$ increases.
  We find
\begin{equation}
k_{z2}^2 = -{\bf k}^2 + 
{\omega^2[\omega^2-({\bf k}\cdot\tilde{\bf v}_A)^2{c}_{s2}^2/c^2]
\over 
\omega^2 \tilde{c}_f^2
-({\bf k }\cdot{\tilde{\bf v}}_A)^2{c}_{s2}^2 }~,
\end{equation}
where $\tilde{v}_A$ is given by equation (8),
$\tilde{c}_s$ is given by equation (10), 
and $\tilde{c}_f^2 \equiv \tilde{c}_{s2}^2+\tilde{v}_A^2$
is the fast magnetosonic wave speed.
   Note that equation (27) with
${\bf K}^2 ={\bf k}^2+k_z^2$ is identical to equation (9)
as it should be.
  Dividing this equation by ${\bf k}^2$ and taking
the square root gives
\begin{equation}
\tilde{k}_{z2}(u) =\pm \left(
{u^2\big[u^2- [\cos(\varphi)\tilde{v}_A]^2{c}_{s2}^2/c^2\big]
\over
u^2 \tilde{c}_f^2
-[\cos(\varphi){\tilde{ v}}_A]^2{c}_{s2}^2 }
-1\right)^{1/2}~.
\end{equation}
   The choice of the sign in this expression is
determined by the condition $\Im(\tilde{k}_{z2}) < 0$.

\subsection{Fundamental Dispersion Relation}

   Equation (16) can be rewritten as
\begin{equation}
{\partial \over \partial z} \delta \hat{p}
=-\big[W\Delta \omega -({\bf k \cdot v}_A)^2\big]\rho~ \delta z~,
\end{equation}
where the terms on the right-hand-side are finite.
Thus we have $\delta\hat{p}(z=\epsilon^+) = 
\delta \hat{p}(z=\epsilon^-)$.
   Using the fact that $\delta \hat{p}_{1,2}
\propto \exp(ik_{z1,2}z)$, we find
$ik_{z2}\delta \hat{p}=-[W\Delta\omega -({\bf k\cdot v}_A)^2]
\rho_2\delta z_2$ and
$ik_{z1}\delta \hat{p}=-\Delta\omega^2 \rho_1\delta z_1$. 
Taking the ratio of these equations gives
$k_{z2}/k_{z1}=(\rho_2/\rho_1)
[\omega^2(1+v_A^2/c^2)-({\bf k \cdot v}_A)^2]/(\Delta
\omega)^2$.
This gives
\begin{equation}
{\rho_1 \over k_{z1}}(\omega -k_x v_x)^2
={\rho_2 \over k_{z2}}\left(1+{v_A^2\over c^2}\right)
\left[\omega^2 -
({\bf k} \cdot\tilde{\bf v}_A)^2 \right]~.
\end{equation}
This is the fundamental dispersion 
relation.  
   It agrees with the result
of HWL in the limit $c\rightarrow \infty$.
   Equation (30) can be rewritten in terms
 of $u=\omega/|{\bf k}|$  as
\begin{equation}
F(u)\equiv \big[u  -\sin(\varphi) v_x\big]^2 -
g^2 \left[{\tilde{k}_{z1}(u) \over \tilde{k}_{z2}(u)}\right]
\big[u^2 -
[\cos(\varphi)\tilde{v}_A]^2 \big]=0~,
\end{equation}
where   $\tilde{k}_{z1}(u)$ is given by equation (26)
and $\tilde{k}_{z2}(u)$ by equation (28), and
\begin{equation}
g^2 \equiv  {\rho_{2} \over \rho_{1}}+
 { {\bf B}^2 \over 4\pi \rho_{1}c^2}~.
\end{equation}
The choice of signs for $\tilde{k}_{z1}$ (equation 26)
and $\tilde{k}_{z2}$ (equation 28) is fixed by the
above mentioned
requirements that $\Im(\tilde{k}_{z1}) >0$ and
$\Im(\tilde{k}_{z2}) < 0$.

  The pressure balance across the interface gives
\begin{equation}
{\rho_1 c_1^2 \over \gamma}
={{\bf B}^2 \over 8 \pi}
+ {\rho_2 c_2^2 \over \gamma}~,
\end{equation}
where $c_2$ is the sound speed in the magnetosphere.
This is the same as
\begin{equation}
 v_A^2 +{2c_2^2 \over \gamma}= 
{\rho_1 \over \rho_2}
{2 c_1^2 \over \gamma}~.
\end{equation}
  Hence  equation (32) can be written as
\begin{equation}
g^2 = {\rho_2 \over \rho_1}\left(1-{2 c_2^2 \over \gamma c^2}\right)
+{2c_1^2\over\gamma c^2} ~~~=~ {\rho_2 \over \rho_1}~,
\end{equation}
where the last equality takes into account that the plasma
motion is assumed to be non-relativistic.

\subsection{Approximate Instability Solution for $g^2 \ll 1$}

  We are interested in conditions where $g^2  \ll 1$ where an
approximate solution to equation (31) can be developed
as follows.    The two terms of equation (31) are written as
\begin{equation}
F=F_0(u) + g^2 F_1(u) =0~.
\end{equation}
We develop a perturbation expansion for this equation based
on the small parameter $g^2$.  
     Thus, we take $u=u_0 +\delta u$ with
$u_0$ is chosen to give $F_0(u_0)=0$ and
 $|\delta u| \ll |u_0|$ assumed.
    Hence
$$
0 \approx F_0(u_0+\delta u)+ g^2 F_1(u_0+\delta u)~,\quad\quad\quad
$$
\begin{equation}
 0 \approx  F_0(u_0)+{dF_0\over du_0}\delta u
+{1\over 2}{d^2F_0 \over d u_0^2}(\delta u)^2+
g^2F_1(u_0) +{\cal O}(g^2|\delta u|)~.
\end{equation}
   We choose $u_0$ such that $F_0(u_0) =0$ which
implies that $u_0 =\sin(\varphi) v_x$.    Then we
have $dF_0/du_0 =0$ and $d^2F_0/du_0^2 =2$.
   Equation (37) then gives   
\begin{equation}
 \delta u = \pm g \left[{\tilde{k}_{z1}(u_0)
 \over \tilde{k}_{z2}(u_0)}\right]^{1/2}\big[\sin^2(\varphi) v_x^2 -
 \cos^2(\varphi) \tilde{v}^2_A \big]^{1/2}~.
 \end{equation} 
 
 For $u_0=\sin(\varphi) v_x$, equation (26) 
 gives $\tilde{k}_{z1}(u_0) = \pm i  $.
    Also, for $u_0=\sin(\varphi) v_x$ and
for $v_x^2 \gg c_{s2}^2$ and $v_x^2 \ll \tilde{c}_f^2$,
equation (28) gives $\tilde{k}_{z2}(u_0) =\mp i$.
   The mentioned conditions on the 
imaginary parts of the $k_{z}$'s then
implies that $\tilde{k}_{z1}/\tilde{k}_{z2} =-1$.
     Therefore, equation (38) implies  instability
for $v_x >\tilde{v}_A (\tan \varphi)^{-1}$ which is
the condition for the Kelvin-Helmholtz 
instability for the considered equilibrium.  
     The real part $\Re(u)=\Re(\omega)/k$
corresponds to the $x-$component of the phase velocity of the
perturbation matching flow speed of 
the external medium $v_x$.    

      For $u_0=\sin(\varphi) v_x$ and $\delta u$ given
by equation (38), one has in region 1,  
$\rho \delta v_{z1}=k_x \delta p_1/ \omega_i$
from equation (14).
    From the equation following equation (11),
we have $\delta z_1 =\delta v_{z1}(\epsilon^+)/\omega_i$,
where $\delta z_1$ is the displacement of the interface.
Consequently, $\delta \rho_1(\epsilon^+)/\rho_1
=\omega_i^2\delta z_1/(k_x c_{s1}^2)$, which
shows that the perturbation in region 1 involves
a change in the density.   The perturbation is
a sound wave evanescent in the $z-$direction with amplitude 
$\propto \exp(i{\bf k\cdot x}-|k_z|z -i\omega_r t)$
where $\omega_r=k_x v_x$ 
and $|k_z|=| k_x|[1+(g v_x/c_{s1})^2]^{1/2}$ from
equation (25).
     The instability results from the interaction of this
wave with the Alfv\'en wave (\S 2.3)
in region 2  which has $\delta \rho_2 =0$ and does not
change the magnitude of the magnetic field.
    Perturbations which change the magnitude
of ${\bf B}$ are suppressed because they
increase the magnetic energy of the system.   
   For this reason the slow and fast magnetosonic
waves are not excited.

  Having the wavevector along the magnetic
field corresponds to bending the field line which
requires energy and is stabilizing.
    Thus the
maximum growth rate occurs
for $\varphi =90^\circ$ (where $k=|k_x|$) 
with no field line bending and is
\begin{equation}
{\rm max}(\omega_i)=\Im(k\delta u) =gk v_x = 
\left({\rho_2 \over \rho_1}\right)^{1/2} k v_x~.
\end{equation}
   This formula for $\omega_i$ applies only for
a restricted range of  $k$:  for say $k > 3  r_m^{-1}$ where
the planar description of the interface is valid, and
for $k < (\Delta r_m)^{-1}$,  where $\Delta r_m$ is
the radial thickness of the interface.     
    The maximum growth rate does not depend 
explicitly on the value of the magnetic field.   
  However, it depends implicitly on $B$ since
the field allows conditions with $\rho_2/\rho_1 \ll 1$
and the field enters the expression for $\tilde{k}_{z2}(u)$. 
   The same formula for the growth rate is found
by Li \& Narayan (2004) who assume incompressible
fluid motion in both media.  
   In our treatment the response in the low-density
magnetized region 2 is an Alfv\'en wave which is
incompressible, but the response in the high-density
unmagnetized region 1 is incompressible {\it only}
in the zeroth approximation where $g=0$ in 
equation (36) and $F_0(u)=0$.   In the first
approximation including  the $g^2$ term the
medium is compressible.  As discussed in \S 4
the compressibility of the region 1 medium gives
rise to observable fluctuations in the emissions
from the interface.

   For comparison with the results of
Arons \& Lea (1980) who took into
account the displacement current
(important for $v_A/c \gtrsim 1$), we
need to give the correspondence of
our variables with theirs.    
    We find that our $\rho_2/\rho_1$ corresponds
with their $\rho_e/\rho_b \ll 1$, that our $v_A$ 
is the same as theirs, $v_A=|{\bf B}|(4\pi \rho_e)^{-1/2}
=|{\bf B}|(4\pi \rho_2)^{-1/2} \gtrless 1$, 
and that their $a= v_A(\rho_e/\rho_b)^{1/2}\ll v_A$. 
    In their equation (A48) we can
neglect  terms involving $\rho_e/\rho_b$ in comparison
with unity.    In the limit where $a^2 \ll 1$,
their equation gives our equation (38) {\it multiplied}
by $(1+v_A^2/c^2)^{1/2}$.    Thus the two
results agree {\it only} in the limit $v_A^2/c^2 \ll 1$.  
  For $a \gtrsim 1$ the two expressions are also
different.

   As a numerical example consider $c_{s1}=c_{s2}=0.0391$,
$v_A=  1.36$, $\tilde{v}_A =0.805$, $v_x=0.167$
(all in units of $c$),  $g^2 =\rho_2/\rho_1 =0.00283$ and
$g=0.0532$, Mach number ${\cal M}_1=v_x/c_{s1} =4.26$,
where we find $\omega_i =0.00887k c$ for
propagation in the $x-$direction.  
   There is instability
for $\varphi> \arctan(\tilde{v}_A/v_x) =78.3^\circ$ up 
to $\varphi=90^\circ$.      
    The range propagation directions of the unstable
waves  is shown in Figure 5 as marked KH.
    The saturation of the exponential growth is
discussed in \S 4.

     We can express equation (39) as
$\omega_i =(\rho_2/\rho_1)^{1/2} k_x c_{s1} {\cal M}_1$
where ${\cal M}_1$ is the Mach number of the external
plasma flow.     In the commonly considered
case of Kelvin-Helmholtz instability between equal
density media  there is instability for waves propagating
parallel to the flow only for ${\cal M} < 2^{3/2}$
(e.g., Hunter \& Whitaker 1989).  
    In the present case there is no similar limit on the
Mach number for the growth of  waves propagating parallel
to the flow.    We of course have the limit $v_x^2 \ll c^2$
because of our assumption of non-relativistic fluid motion.

\section{Nonlinear Effect of Unstable KH Modes}

   A large number of computer simulation studies 
have been done  on the nonlinear
evolution of the KH instability for
different initial configurations (e.g.,  Zhang, MacFadyen, \&
Wang 2009; Keppens et al. 1999; Frank et al. 1996).
The studies do not address the configuration considered here,
but they suggest the  approximate treatment discussed below.
A comparison is made with the nonlinear model of Burnard et al. (1983).

    Starting from a sharp interface,  
exponential growth of the KH ceases at
some time when typically one or more
 ``Kelvin's Cat Eyes''  form.  Subsequently
the interiors of the Eyes may become 
highly irregular but their widths remain 
roughly constant.
   For simplicity, we consider the region of the
magnetopause  outside of $r_m$ or equivalently
$z>0$ (region 1).
    Also we assume that the wave propagates
in the $x-$direction where the growth rate is
a maximum.   
       From \S 3 it is clear that 
the unstable wave initially corrugates the interface 
displacing it by the amount 
$\delta z = i\delta v_{z1}(\epsilon^+)/\Delta \omega$.
   The exponential growth will cease when the
width of the Cat Eye is of the order of the reduced
wavelength.  That is, 
$|\delta z| = q\lambdabar$,  where $q=$const is a number
of the order unity and $\lambdabar = 1/k$.
     At later times the initially sharp interface is
effectively smoothed  over a distance $\sim \delta z$
giving the interface a radial thickness $\Delta r_m =\delta z$
(Frank et al. 1996).    This finite thickness  acts
to stabilize the KH instability for a given $k=\lambdabar^{-1}$.
    At saturation $|\delta v_{z1}|= q\omega_i/k$
since $\Delta \omega =i\omega_i$.   
    From equation (14)  we have $\rho_1\delta v_z 
= i |k| \delta p/\Delta \omega$.  
    This gives $\delta p/\rho_1 =
q(\omega_i/k)^2$ and $\delta \rho /\rho_1 =q\omega_i^2/(k c_{s1})^2
=qg^2 (v_x/c_{s1})^2$ at saturation.
     The use of the linear relations between the
fluid variables is plausible in the nonlinear regime
if   $|\delta \rho|/\rho_1 =q g^2(v_x/c_{s1})^2 \ll  1$.     

     In general, there is a spectrum of saturated KH waves with
 the contribution to $|\delta \rho |^2$  from wavenumbers
 of order $k$ equal to  $k |\delta \rho_k |^2$, where
 $|\delta \rho_k|^2$ is the wavenumber power spectrum of
 $\delta \rho$.   
     From the previous paragraph we have
 \begin{equation}
\big|\delta \rho_k\big|^2={ \rho_1^2 q^2 g^4 \over k}
  \left({v_{x} \over c_{s1}}\right)^4~.
\end{equation}  
The total mean-squared density fluctuation is therefore
$\langle|\delta \rho|^2\rangle=\int dk  |\delta \rho_k|^2=
\rho_1^2 q^2g^4 (v_{x}/c_{s1})^4 
\ln(k_{\rm max}/k_{\rm min})$.
Here, $k_{\rm min}$ is $2\pi$ over the longest
wavelength  which is of the order of the size  of the
patch of the interface ($\ll r_m$), and $k_{\rm max}$
is $2\pi$ over the shortest wavelength which may be
the radial thickness of the magnetopause discussed
further below.   Equation (40) corresponds to a 
``pink-noise'' or ``one-over-$f$'' spectrum.   
      For the assumed  isentropic equation of state  there
are corresponding temperature fluctuations  
$|\delta T_k|^2/T^2 = (\gamma-1)^2|\delta \rho_k|^2/\rho^2$
which can give  spatial variations in the radiation
from the optically thin regions of the magnetopause. 
     Connecting the spectrum of waves with observed
noise spectra of X-ray sources (van der Klis 2006)
is, however, beyond the scope of this work since it
involves integration over the entire magnetopause
as well a treatment of the radiation transfer.
  
      Figure 6 shows a sample realization of the interface
with the wavenumber spectrum given by Equation (40)
in a reference frame comoving with the disc plasma.

\begin{figure}
\centering
\includegraphics[scale=0.5]{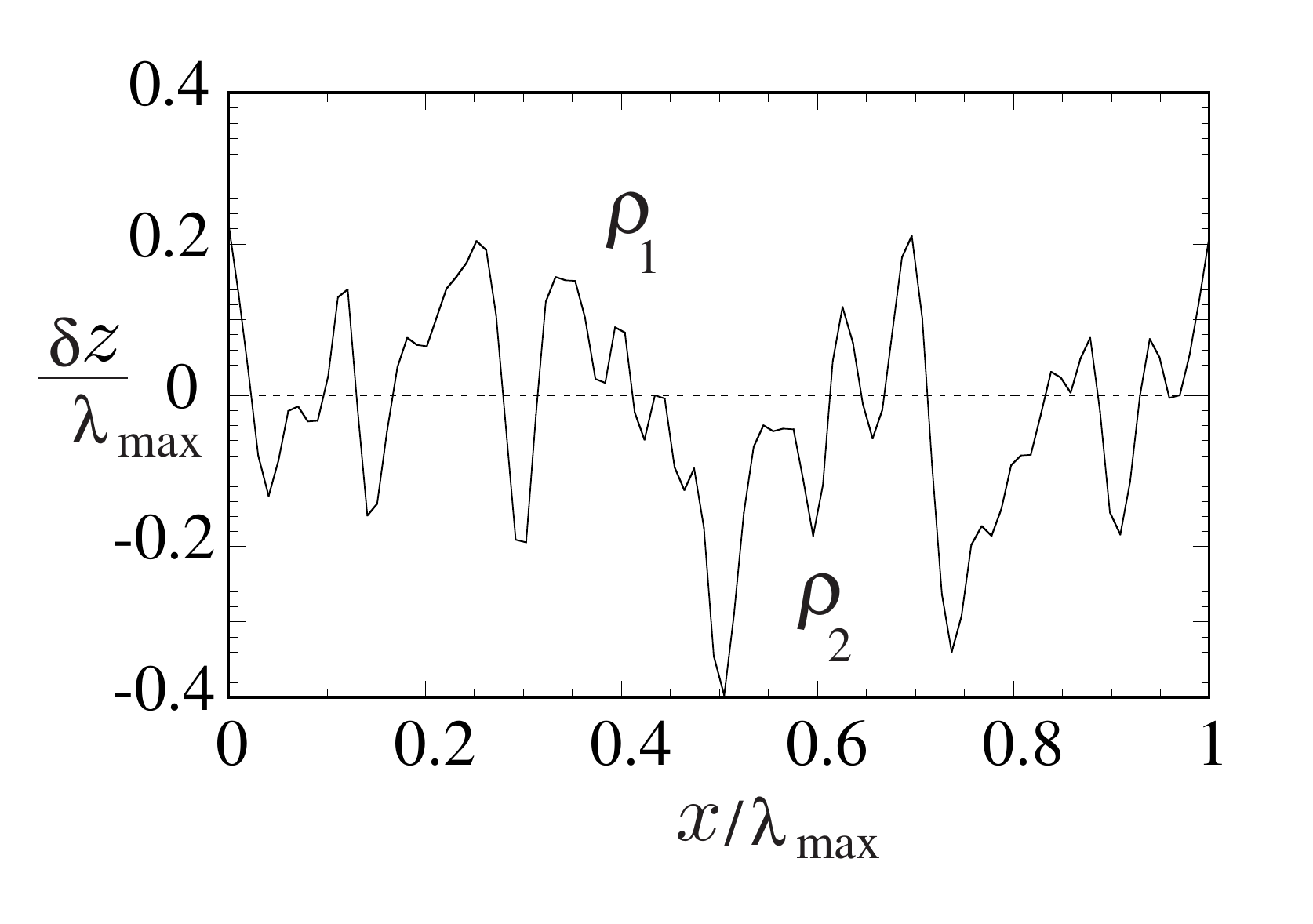}
\caption{Sample realization of the saturated interface
for $q=1$ and $k_{\rm max}/k_{\rm min}=50$
viewed in the reference frame comoving with 
the disc plasma ($\rho_1$).
The Fourier amplitudes $\delta \rho_k$ are generated with
a random phase and magnitude with rms magnitude
$\propto k^{-1/2}$.
}
\end{figure}

    The equilibrium flow of \S 2.1 does not include
accretion.  However, the unstable KH modes of    
\S 3.2 can give rise to accretion in the vicinity of 
the magnetopause radius $r_m$.    
   The mass flux per unit $(x,y)$ area ($A$) in the
$z-$direction    across the
magnetopause is simply
\begin{equation}
{\delta \dot{M}_{KH} \over \delta A}
= \Re \big(\delta \rho \delta v_z^*\big)~,
\end{equation}
where $\delta \rho =\rho(k_x \delta v_z +k_z \delta v_z)/\Delta \omega$
from equation (13), $\rho \delta v_x =k_x \delta p/\Delta \omega$,
and $\rho \delta v_z =i |k_z|\delta p/\Delta \omega$ from
equations (14).
 Substitution gives
 \begin{eqnarray}  
 {\delta \dot{M}_{KH} \over \delta A}
& = &~-~ {2 k^3c_{s1}^2 \langle\big|\delta 
\rho\big|^2\rangle \over \rho_1 \omega_i^3}
\nonumber\\
&=& ~-~2q^2g \rho_1 v_x \ln\left({k_{\rm max}
\over k_{\rm min}}\right) ~,
\end{eqnarray}
where we have used equations (23) and (39).

      Equation (42) corresponds to an  effective
{accretion speed} at the magnetopause of $u^{\rm accr}_{KH}=
2q^2g v_x \ln(k_{\rm max}/k_{\rm min})$.  
    The accretion speed proposed by Burnard et al. (1983)
agrees with $u^{\rm accr}_{KH}$ if $q = 0.11$ for     
$k_{\rm max}/k_{\rm min}=50$.
      For the values given below equation (39) and $q=0.1$,
the accretion speed is $0.00069c$.  
For comparison, in a standard alpha-disc model
(Shakura \& Sunyaev 1973) the accretion speed
is of the order of $u_{\rm disc}=\alpha(c_s/v_K)^2 v_K$,
where $v_K$ os the Keplerian velocity.
   We find
$u_{\rm disc}\approx 10^{-3}c$ assuming $\alpha=0.1$, $c_s=c_{s1}$,
and $v_K =v_x$.
    On the other hand for
conditions where $q={\cal O}(1)$ and $u^{\rm accr}_{KH} > u_{\rm disc}$,
we suggest that the profiles of the density and other
variables (shown in Figure 1) become less steep having
a width $\Delta r_m$.  
    The present treatment of the KH instability 
remains valid for reduced wavelengths $\lambdabar > \Delta r_m$
which corresponds to $k <k_{\rm max} =(\Delta r_m)^{-1}$.   
    Clearly, for less steep profiles the  factor   
$\ln(k_{\rm max})/k_{\rm min})$ decreases as does
$u^{\rm accr}_{KH}$.   Thus  self-consistent conditions
can evolve naturally to give $u^{\rm accr}_{KH}=u_{\rm disc}$.

    The unstable KH waves also gives a radial influx  of
angular momentum (about the $z-$axis) per unit area  of
the magnetospheric patch.   
For simplicity we neglect the rotation of
the magnetosphere, $(\Omega_* r_m)^2 \ll v_K^2(r_m)$.
Then,
\begin{eqnarray}
{\delta {F}_{KH}\over \delta A} &=&r_m \big[
\Re(\delta \rho v_x \delta v_z^*)
+ \Re(\rho \delta v_x  \delta v_z^*)\big]~,
\nonumber\\
&=& r_m v_x{\delta \dot{M}_{KH} \over \delta A}
\end{eqnarray}
where the term $\Re(\rho \delta v_x \delta v_z)$ vanishes and
where $r_m v_x$ is the specific angular momentum of
the external disc matter outside the magnetopause.

\section{General orientation of the magnetic field}

    For the general case 
 the star's magnetic
moment $\rvecmu$ is not aligned with $\rvecOmega_*$.
However, we assume the rotation axis of the disc is aligned
with $\rvecOmega_*$.    The shear layer between the
disc and star will be inherently time dependent due
to the star's rotation.   
    For example, for an orthogonal rotator where
$\rvecmu$ is in the equatorial plane, the plasma
in the disc sees a magnetic field reversing its
direction with an angular frequency $\omega_B=v_x/r_m =
(v_K-r\Omega_*)_{r=r_m}$.    This frequency
may be larger or smaller than the growth rate
of the KH instability.  
   From equation (39) we have  $\omega_i/\omega_B=
g (r_m/\lambdabar)$.  
     For  $\omega_i/\omega_B >1$
significant wave growth can occur before the moves
from a given field region.  
    In the other limit the wave growth may be
recurrent as a given plasma region returns to
a given field region.    
   In the following we consider $\bf B$ to
be time-independent.

   For a general field orientation,
$\bf B$ is still in the plane of the shear
layer, that is, the $(x,y)$ plane of Figure 2.  
An equilibrium with a $B_z$ component is not
possible.     Note however that the $z-$direction is not
necessarily in the $\hat{\bf r}-$direction.
     For this case we discuss
the needed modification of  Figure 5 where
it is appropriate to use the angle $\phi$ rather 
than $\varphi$.
The flow
velocity remains in the $x-$direction so
that the fast and slow sound wave curves
remain unchanged as does the dashed circle.
  The dashed circle can be written as
$v_x\cos(\phi)$ for $-\pi/2 \leq \phi \leq \phi$.  
   What changes in Figure 5 is that the ``figure eight'' curve for
the Alfv\'en wave is  rotated by angle 
 $\theta$ say in the counter-clockwise direction.
   The figure-eight curve is give by $\tilde{v}_A| \sin(\phi -\theta)|$.
      In place of equation (38) we find
\begin{equation}
 \delta u = \pm g \left[{\tilde{k}_{z1}(u_0)
 \over \tilde{k}_{z2}(u_0)}\right]^{1/2}\big[\cos^2(\phi) v_x^2 -
 \sin^2(\phi-\theta) \tilde{v}^2_A \big]^{1/2}~.
 \end{equation}  
 The limits on the directions of the unstable Kelvin-Helmholtz
modes are given again by the intersection of the dashed
circle in Figure 5 and the rotated figure-eight curve.
These limits are easily found to be
\begin{equation} 
\phi_{1,2} = \arctan\left( \tan(\theta) \pm {v_x \over
\tilde{v}_A |\cos(\theta)|}\right)~.
\end{equation}  
Figure 7 shows the dependence of the real and
imaginary parts of the KH most unstable mode on
the tilt angle of the field $\theta$.

\begin{figure}
\centering
\includegraphics[scale=0.5]{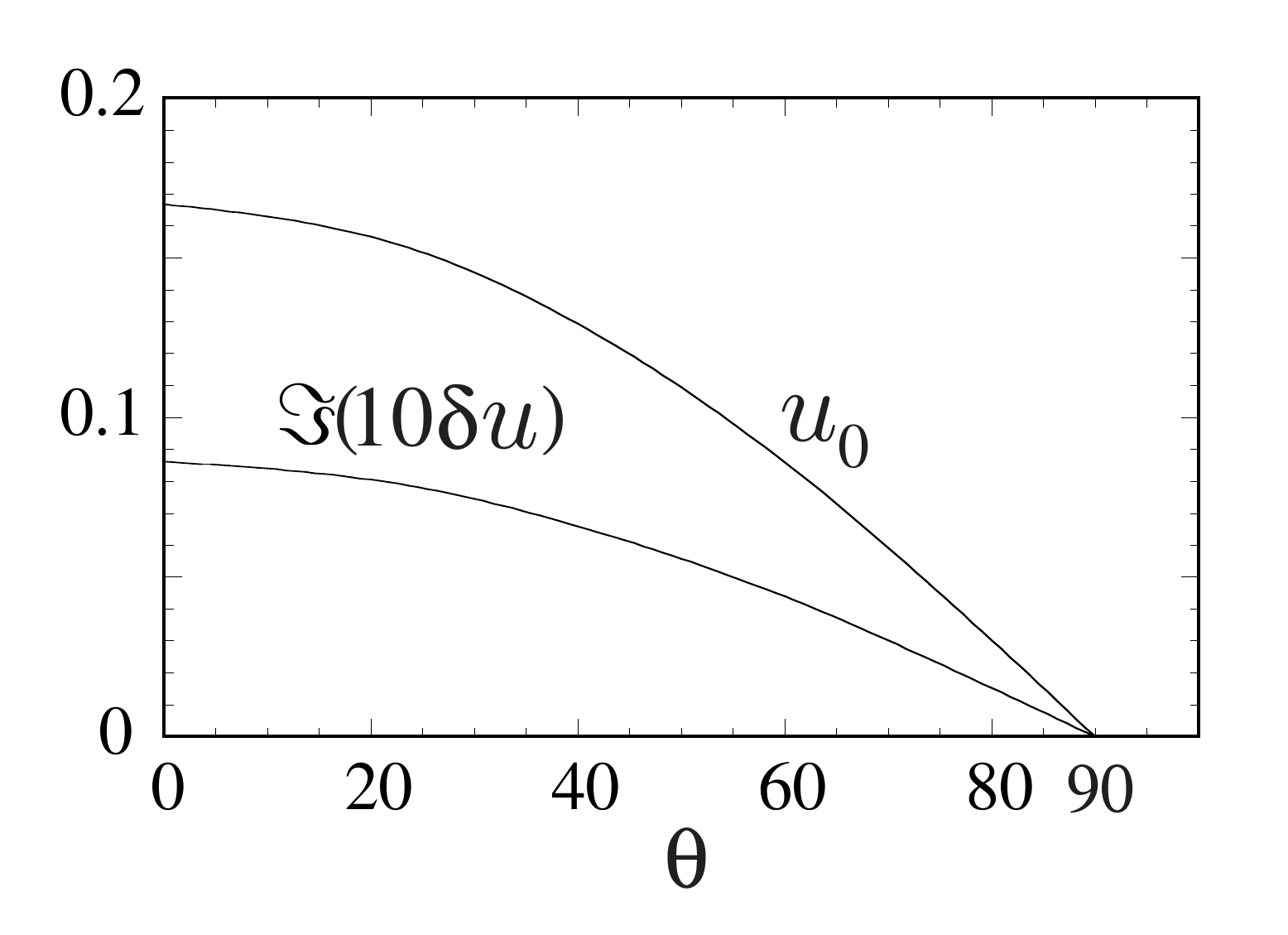}
\caption{Dependences of the real part of the phase
velocity $u_0 =\omega_r/k$ and the imaginary part
$\Im(\delta u)= \omega_i/k$ (in units of $c$) 
on the tilt angle of the
magnetic $\theta$ for the most unstable wave.
   The parameters are the same
as for Figure 5. 
}
\end{figure}

\section{Conclusions}

     This work investigated the short wavelength ($\lambda \ll 2\pi r_m$)
stability of the magnetopause at radius $r_m$  between  
a rapidly-rotating, supersonic, dense ($\rho_2$) accretion disc and a slowly-rotating
low-density magnetosphere ($\rho_1 \gg \rho_1$) of a magnetized star.
   The magnetopause is  a strong shear layer
with  rapid changes in the azimuthal velocity, the density,
and the magnetic field over
a short radial distance ($\Delta r_m \ll r_m$),
 and thus the Kelvin-Helmholtz (KH) instability   
may be important.    
       This work has focused on the case of a star with an aligned
dipole magnetic field so that the magnetic field
is axial in the disc midplane  and
perpendicular to the disc flow velocity.   
      For the aligned dipole case the magnetopause is most  unstable     
for KH waves propagating  in the azimuthal
direction perpendicular to the magnetic field.   
     Propagation not perpendicular to the magnetic field 
changes the magnetic field which gives a stabilizing effect.  
The growth rate of the instability is    $\omega_i =
(\rho_2/\rho_1)^{1/2} k_x v_x$, where $v_x$ is 
the velocity of the disc plasma relative to that of
the magnetospheric plasma which corotates with the star.
    The wave phase velocity is that of the disc matter.

    We discussed the non-linear saturation of the instability
which we argued occurs for a mode  of 
 wavelength $\lambda=2\pi/k$ when the displacement of
the interface between the disc and magnetospheric
plasmas is of the order of $\lambda/2\pi$. 
      From this we developed a quasi-linear model which 
led to a wavenumber  power spectrum $\propto k^{-1}$
of the density and temperature fluctuations of the
magnetopause. 
    The quasi-linear model gave  the mass accretion 
and angular momentum inflow rates across the
magnetopause.       
     The mass accretion rate
(per unit area of the magnetopause)
was found to be $2(\rho_2/\rho_1)^{1/2} \rho_1 v_x
\ln(k_{\rm max}/k_{\rm min})$, where $k_{\rm max, min}$
are the maximum and minimum wavenumbers discussed
in \S 4.
    For self-consistent conditions
this mass accretion rate will be equal to the 
disc accretion rate at large distances from
the magnetopause.

    We also considered the case where the
magnetic field is not perpendicular to the
flow velocity but tilted by an angle $\theta$
relative to the $z-$axis, where $\theta=0$
corresponds to the aligned rotator case
treated earlier.   We found that the maximum
growth rate and the associated wave
phase velocity decrease monotonically
with $\theta$ and that both approach 
zero as $\theta \rightarrow 90^\circ$.
That the growth rate goes to zero
is a result of the stabilizing effect
of the magnetic field.  
      Thus  an
orthogonal rotator where the
star's magnetic moment $\rvecmu$
is perpendicular to the star's
rotation axis $\rvecOmega_*$ is
stable to the KH mode.

\section*{Acknowledgements}

We thank Prof. David Chernoff for valuable discussions on
this topic which stimulated the initiation of this work.
The authors (RVEL and MMR) were supported in 
part by NASA grant NNX08AH25G and by
NSF grants AST-0607135 and AST-0807129.


\begin{thebibliography}{}

\bibitem{} Arons, J., \& Lea, S.M. 1980, ApJ, 235, 1016

\bibitem{} Burnard, D.J., Lea, S.M., \& Arons, J. 1983, ApJ, 266, 175
 
\bibitem{} Chandrasekhar, S. 1961, {\it Hydrodynamic
and Hydromagnetic Stability} (Oxford Press: London),
p. 481


\bibitem{} Faganello, M., Califano, F., \& Pegoraro, F. 2008, PRL,
101, 175003

\bibitem{} Frank, A., Jones, T.W., Ryu, D., \& Gaalaas, J.B. 1996,
ApJ, 460, 777

\bibitem{} Hardee, P.E. 2007, ApJ, 664, 26

\bibitem{} Hunter, J.H., \& Whitaker, R.W. 1989, ApJS, 71, 777

\bibitem{} Hunter, J.H., Whitaker, R.W., \&
Lovelace, R.V.E. 1998, ApJ, 508, 680



\bibitem{} Jackson, J.D. 1975, {\it Classical
Electrodynamics}, Second Edition (John Wiley \& Sons:
New York), p. 489

\bibitem{} Keppens, R., T\.oth, Westermann, R.H.J., \&
Goedbloed, J.P.  1999, J. Plasma Physics, 61, 1

\bibitem{} Kulkarni, A.K., \& Romanova, M.M. 2008, MNRAS, 386, 673

\bibitem{} Landau, L.D., \& Lifshitz, E.M. 1987, {\it Fluid Mechanics},
(Pergamon Press: Oxford), p. 315).

\bibitem{} Li, L.-X., \& Narayan, R. 2004, ApJ, 601, 414

 \bibitem{} Lovelace, R.V.E., Li, H., Colgate, S.A., \&
Nelson, A.F. 1999, ApJ, 513, 805

\bibitem{} Lovelace, R.V.E., \& Romanova, M.M. 2007, ApJ,
670, L13 (LR07)

\bibitem{} Lovelace, R.V.E., Turner, L., \& Romanova, M.M. 2009,
ApJ, in press (arXiv:0905.1071) 

\bibitem{} Miura, A., \& Pritchett, P.L. 1982, JGR, 87, 7431

\bibitem{} Osmanov, Z. Mignoone, A., Massaglia, S., Bodo, G.,
\& Ferrari, A. 2008, A\&A, 490, 493

\bibitem{} Romanova, M.M., Kulkarni, A.K., \& Lovelace,
R.V.E. 2008, ApJ, 673, L171

\bibitem{} Roy-Choudhury, S., \& Lovelace, R.V.E. 1986,
ApJ, 302, 188

\bibitem{} Shakura, N.I., \& Sunyaev, R.A. 1973, A\&A, 24, 337

\bibitem{} Tsang, D., \& Lai, D. 2009, MNRAS, 396, 589

\bibitem{} van der Klis, M. 2006, in {\it Compact Stellar
X-Ray Sources}, Eds. W.H.G. Lewin \& M. van der Klis (Cambridge:
Cambridge Univ. Press), p. 39

\bibitem{} Zhang, W., MacFadyen, A., \& Wang, P. 2009, ApJ, 692,
L40

\end{thebibliography}
\end{document}